\documentclass[12pt,english]{article}
\usepackage{mathpazo}
\usepackage[T1]{fontenc}
\usepackage[latin9]{inputenc}
\usepackage{geometry}
\usepackage{babel}
\usepackage{amssymb}
\usepackage[authoryear]{natbib}
\usepackage[unicode=true]
 {hyperref}

\makeatletter
\newcommand{\lyxaddress}[1]{
\par {\raggedright #1
\vspace{1.4em}
\noindent\par}
}

\makeatother

\begin{document}

\title{Reply to Benestad\textquoteright{}s comment on \textquotedblleft{}Discussions
on common errors in analyzing sea level accelerations, solar trends
and global warming\textquotedblright{} by Scafetta (2013)}

\author{Nicola Scafetta$^{1,2}$}

\maketitle

\lyxaddress{$^{1}$Active Cavity Radiometer Irradiance Monitor (ACRIM) Lab, Coronado,
CA 92118, USA }

\lyxaddress{$^{2}$Duke University, Durham, NC 27708, USA}

\lyxaddress{Correspondence to: N. Scafetta (nicola.scafetta@gmail.com)}
\begin{abstract}
Herein I respond to the criticism and to the complains by Benestad
(Pattern Recogn. Phys. 1, 91-92, 2013, \href{http://dx.doi.org/10.5194/prp-1-91-2013}{http://dx.doi.org/10.5194/prp-1-91-2013})
of Scafetta (Pattern Recogn. Phys. 1, 37-57, 2013, \href{http://dx.doi.org/10.5194/prp-1-37-2013}{http://dx.doi.org/10.5194/prp-1-37-2013})
that I found misleading and not tenable. More significantly, Benestad
did not find any physical nor mathematical error in Scafetta's work.
Thus, Scafetta scientific results remain fully confirmed. \medskip{}
\newline Please cite this article as: Scafetta, N.: Reply to Benestad's
comment on ``Discussions on common errors in analyzing sea level
accelerations, solar trends and global warming'' by Scafetta (2013).
\textit{Pattern Recogn. Phys.} 1, 105-106, doi:10.5194/prp-1-105-2013,
2013.\newline \href{http://dx.doi.org/10.5194/prp-1-105-2013}{http://dx.doi.org/10.5194/prp-1-105-2013}
\end{abstract}
I found \citet[hereafter B13]{Benestad2013} criticism of \citet[hereafter S13]{Scafetta2013}
misleading. More significantly, B13 did not find any physical nor
mathematical error in S13. Thus, S13 scientific results remain fully
confirmed. 

B13 complained that S13 would have misrepresented \citet[hereafter BS09]{Benestad2009}
because BS09 too argued that the regression models studied in BS09
are flawed because of multicollinearity of the constructors. However,
BS09 did conclude that ``the most likely contribution from solar
forcing a global warming is $7\pm1\%$ for the 20th century'' using
regression models of the global surface temperature. S13 questioned
this conclusion and demonstrated that their result is consistent only
with outdated \textit{hockey-stick} paleoclimatic temperature reconstructions
\citep[e.g.][]{Mann1999}.

BS09 studied two regression models for the 1900-2000 global mean temperature.
Equation 1 uses as constructors the solar and GHG forcings, which
are collinear because both trend upward: $\left\langle T\right\rangle =\alpha_{0}+\frac{0.7}{4}\alpha_{1}S+5.35\alpha_{2}\times\ln(\rho)+\mu$
\citep[e.g.][]{Lean2008}. Equation 2 uses the 10 GISS forcings, and
9 functions out of 10 are collinear to each other: $\left\langle T\right\rangle =\beta_{0}+\beta_{1}F_{S}+\beta_{2}F_{GHG}+\beta_{3}F_{O3}+\beta_{4}F_{H2O}+\beta_{5}F_{land}+\beta_{6}F_{snow}+\beta_{7}F_{Aer}+\beta_{8}F_{BC}+\beta_{9}F_{Refl}+\beta_{10}F_{AIE}+\mu$.
The solar contribution to the 20th century warming is estimated to
be $\sim10$\% using Eq. 1 and $\sim7$\% using Eq. 2. The latter
result is consistent also with the GISS ModelE prediction \citep[figure 6D]{Scafetta2013}. 

B13 complained that BS09 compared ``climate models and observations''
that ``only included two co-variates,'' not 10 as shown in BS09
Equation 2 and as S13 would have claimed. However, S13 (table 1) demonstrated
that during the 20th century solar forcing is collinear ($r\approx0.7$)
with other 8 constructors taken \textit{singularly}. Volcano forcing
was the only exception. A regression model is misleading also if it
is based on just two collinear constructors, as B13 claimed to have
done. At the end, BS09 ``7\%'' claim is only supported by the GISS
ModelE prediction; the result remained not validated by robust data
analysis and, therefore, BS09 argument falls into circular reasoning.

Indeed, demonstrating the multicollinearity flaws of the regression
models studied in BS09 was not the ultimate goal of S13. S13 proposed
a solution using the regression methodology in a situation where the
solar forcing is not collinear ($|r|\lessapprox0.2$) with other forcings
(S13, table 3). Thus, B13 misunderstood S13 argument that was based
on two major results: (1) S13 (figure 6) demonstrates that GISS ModelE
underestimates the empirical solar signature by a factor varying from
3 to 8; (2) S13 (figure 7) demonstrates that BS09 conclusion that
the sun contributed only $\sim7\%$ of the 20th century warming is
compatible only with outdated \textit{hockey-stick} paleoclimatic
temperature reconstructions. For example, \citet{Mann1999} temperature
reconstruction shows a preindustrial climatic variability of $\sim0.2$
$^{o}C$ that implies a very small climatic solar effect. Yet, more
recent paleoclimatic temperature reconstructions \citep[e.g. ][]{Moberg}
show a far greater preindustrial climate variability ($\sim0.7$ $^{o}C$)
implying a strong climatic solar effect yielding a solar contribution
to the 20th century warming comparable with the anthropogenic one.
These results fully confirm Scafetta and West's earlier works (2005,
2006) that BS09 criticized.

S13 (section 4) demonstrated that BS09 misapplied the Maximum Overlap
Discrete Wavelet Transform (MODWT) by erroneously adopting the periodic
padding instead of the reflection one yielding Gibbs artifacts. B13
acknowledged BS09 math error, but surprisingly complained that \citet{Scafetta2006}
provided insufficient analysis details. Benestad's opinion is not
tenable, however. Using the reflection padding in decomposing trending
sequences (e.g. the global surface temperature and the total solar
irradiance records from 1900 to 2000) is a standard technique of analysis
detailed in \citet{Percival}, which was properly referenced in \citet{Scafetta2006}.
Moreover, S13 (figure 9) demonstrated that using the erroneous periodic
padding yields a severe physical incongruity: the climate cooled when
the solar forcing increased from 1995 to 2000. Noting physical incongruities
to check calculations is standard practice in time series analysis.
Thus, BS09 just misapplied MODWT. Still, B13 claimed that BS09 MODWT
errors would not matter. Yet, B13 claim is contradicted by S13 results
(table 6, figure 9 and 10) and by standard signal processing strategies
aimed to avoid Gibbs artifacts.

B13 argued that ``taking the relative magnitudes between two bandpass
filtered signals, does not identify a true connection between the
two.'' Yet, the ability of identifying specific (solar-induced temperature)
fingerprints depends on the signal-to-noise ratio strength. This strength
is very small in the GISS ModelE simulations used in BS09 to test
Scafetta's criticized methodology, but not in the temperature records
where the methodologies are expected to properly work \citep{Scafetta2009,Scafetta2013b}.
Assessing the skill of methods in situations where they work poorly
and concluding that in no situation can they work, is logically flawed.
In addition, B13 did not acknowledge that Scafetta's attribution was
also based on two supporting considerations: (1) on correlation analyses,
where the temperature oscillations were found sufficiently synchronous
to the correspondent solar oscillations \citep{Scafetta2005,Scafetta2009,Scafetta2013b};
(2) on comparisons with similar empirical results found in the scientific
literature that were obtained with alternative methodologies.

Essentially, B13 argued that attribution results based on data analysis
must be rejected simply because they might be coincidental. Yet, regression
and filtering methodologies are widely used in science. In absence
of math errors, the scientific method requires critics to propose
alternative and superior physical explanations to reject an interpretation
based on data analysis. Alternative physical proposals to explain
the observed climatic oscillations are missing in BS09 and B13. 

B13 (last paragraph) wrongly claimed that S13 ``further made reference
to `outdated hockey-stick paleoclimatic temperature graphs' with no
factual support.'' Yet, S13 did provide the necessary support by
referencing \citet{Moberg}, \citet{Mann2008}, \citet{Ljungqvist}
and \citet{Christiansen} that proposed novel paleoclimatic temperature
reconstructions demonstrating a far greater pre-industrial climate
variability than the original \textit{hockey-stick} temperature graphs
\citep[e.g.][]{Mann1999}. Finally, because most of S13 supporting
references were published \textit{after} the 2007 IPCC assessment
report, B13 argument that in 2007 the IPCC still supported the original
\textit{hockey-stick} temperature reconstructions cannot be cogently
used against S13. Additional critiques of the hockey-stick records
were provided also by \citet{McShane}.

\section*{Appendix}

On the web-site of the journal (Pattern Recognition in Physics) where
\citet{Benestad2013} was published, Benestad linked his paper to
\citet{Benestadet} submitted to Earth System Dynamics (\href{http://www.pattern-recogn-phys.net/1/91/2013/prp-1-91-2013-relations.html}{http://www.pattern-recogn-phys.net/1/91/2013/prp-1-91-2013-relations.html}).
In their paper Benestad and collaborators attempted a refutation of
a number of peer-reviewed published works on several journals authored
by several authors including some authored by Scafetta. However, \citet{Benestadet}
received strong criticisms and rebuttals during its open review process:
see \href{http://www.earth-syst-dynam-discuss.net/4/451/2013/esdd-4-451-2013-discussion.html}{http://www.earth-syst-dynam-discuss.net/4/451/2013/esdd-4-451-2013-discussion.html}.
As a consequence \citet{Benestadet} was rejected by the editor of
Earth System Dynamics. My own 32-page rebuttal is found in \citet{Scafetta2013c}.

\end{document}